\documentclass[12pt]{article}
\usepackage[english]{babel}
\usepackage[utf8]{inputenc}
\usepackage{johd}

\usepackage{float}          
\usepackage{times}       
\usepackage{amsmath,amsfonts,amssymb,amscd,amsthm,xspace}         
\usepackage{algorithmic}         
\usepackage{algorithm2e}        
\usepackage{verbatim}        
\usepackage{appendix}         
\usepackage{natbib}
\usepackage{hyperref}
\usepackage{cleveref}
\usepackage{subcaption}

\title{Cutting the Geopolitical Ties: Foreign Exchange Reserves, GDP and Military Spending}

\author{Boris Podobnik$^{a,b,c}$, Dorian Wild$^{b}$, Dejan Kovac$^{d}$  
}

\date{May 2025} 

\begin{document}
\maketitle
\thispagestyle{empty}

\vspace{25pt}

\begin{abstract} 
\noindent We show that the amount of foreign exchange reserves (FER) in the world in a given currency is highly correlated with the GDP and military spending of that country for a set of western economies during the last 20 years. Taking into account multicollinearity,  Ridge and Lasso regressions reveal that the Foreign Exchange Reserve is better explained by military spending than GDP for seven western currencies.  For each year shown, military spending is statistically significant more than the monetary instrument $M2$. Comparing the currency of the second world economy, the Chinese renminbi, is well beyond the western FER equilibrium, but yearly analysis shows that there is a steady trend towards a new FER balance. Next, we define a complex geopolitical network model in which the probability of switching to an alternative FER currency depends both on economic and political factors. Military spending is introduced into the model as an average share of GDP observed within the data. As the GDP of a particular country grows, so does the military power of a country. The nature of the creation of new currency networks initially depends only on geopolitical allegiance. As the volume of trade with a particular country changes over a designated threshold, a country switches to the currency of that country due to increased trade. If the current steady trend continues within the same geopolitical setting as in the past twenty years, we extrapolate that the RMB and Western currencies could reach a new FER balance within 15 to 40 years, depending on the model setup.

\end{abstract}

\noindent\keywords{foreign exchange reserves, military spending, geopolitics, US, China }\\

\vspace{15pt}

\noindent\small $^{a}$Faculty of Civil Engineering, University of Rijeka, 51000 Rijeka, Croatia, EU 
\\
\small $^{b}$Zagreb School of Economics and Management, 10000 Zagreb, Croatia, EU
\\
\noindent\small $^{c}$Faculty of Informatics,  8\,000 Novo Mesto,  Slovenia, EU
\\
\noindent\small $^{d}$Center for International Development, Harvard Kennedy School, Harvard University

\noindent\small $^{*}$
 The authors thank Rosario N. Mantegna for useful discussions. B.P. received support from the Slovenian Research Agency (ARRS) through Projects J7-3156B.  

\clearpage

\section{Introduction}

\begin{figure}[!hb]
  \centering 
  \caption{China's holdings of U.S. Treasuries }\includegraphics[width=0.9\textwidth]{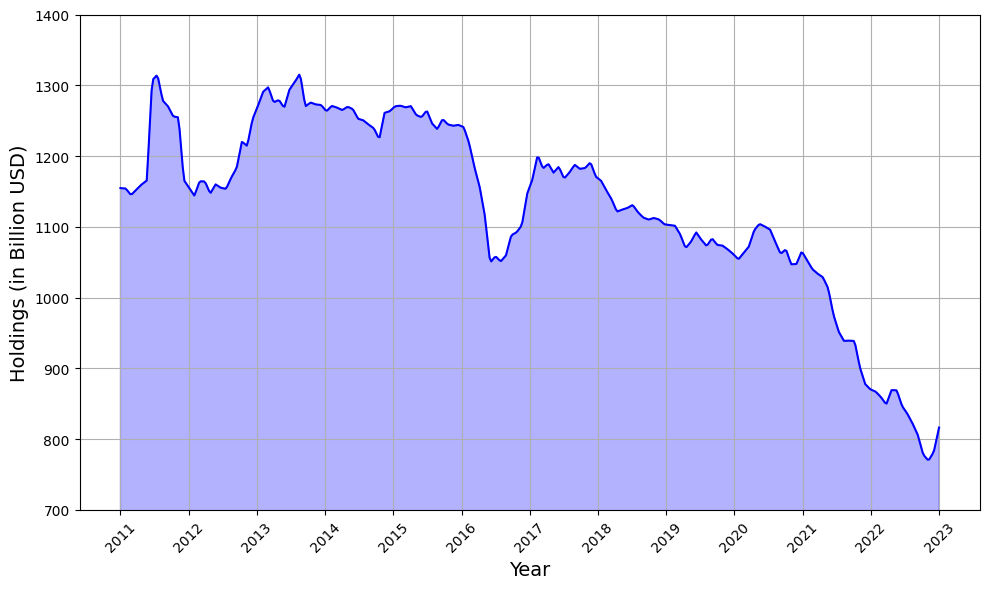}\\
  
  \label{CHNUS}
\end{figure}

\begin{figure}[!hb]
  \centering 
  \caption{Abrupt decline in Russia's holdings of U.S. Treasuries  }\includegraphics[width=0.9\textwidth]{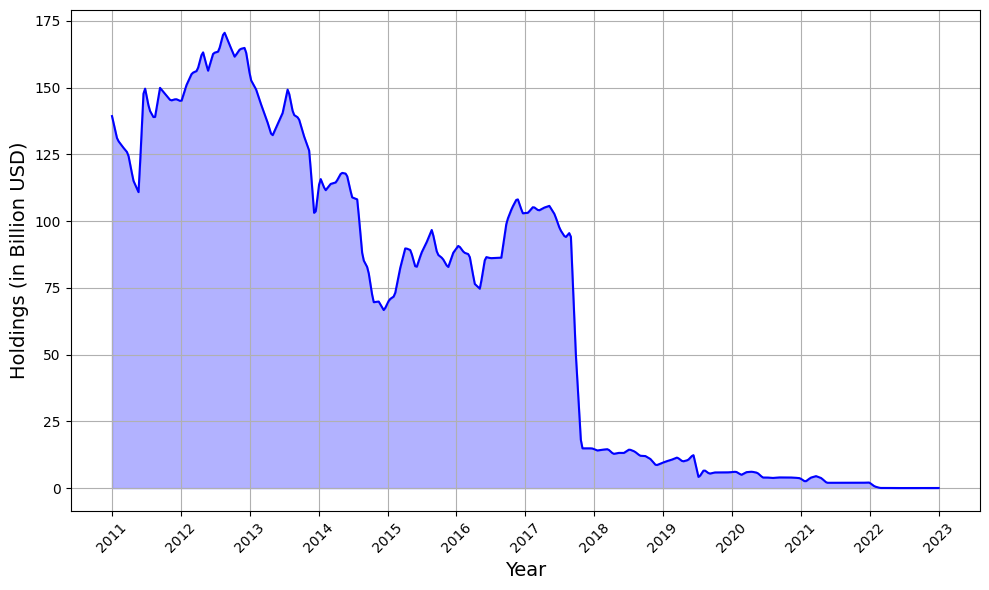}\\
  
  \label{RUSUS}
\end{figure}

\begin{figure}[!hb]
    \caption{Increase in  India's holdings of U.S. Treasuries  }
  \centering \includegraphics[width=0.9\textwidth]{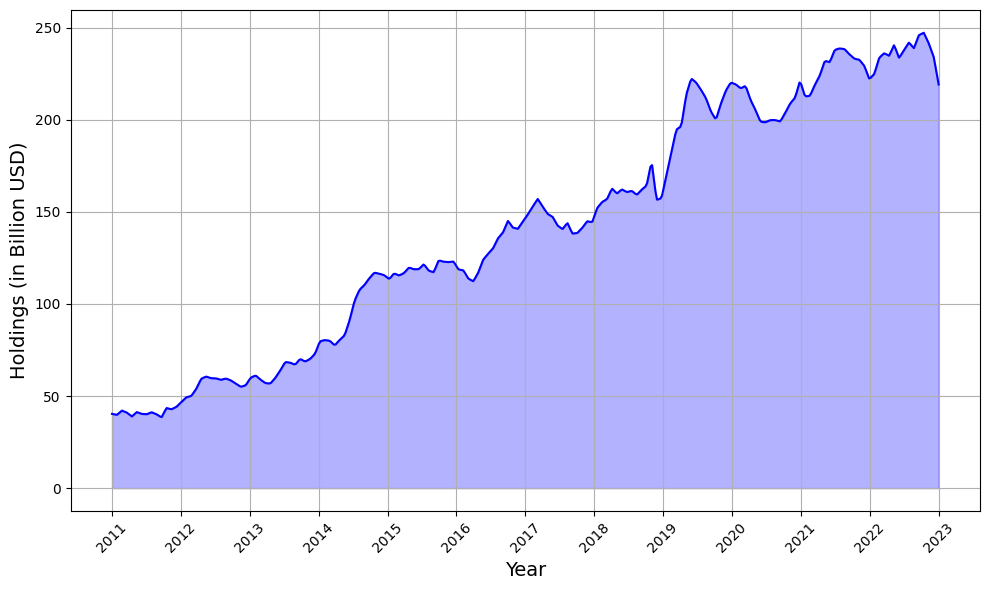}\\
  
  \label{INDUS}
\end{figure}

How do countries decide what basket of currencies to use within their foreign exchange reserves has been a research question that has puzzled scientists in both political science and economics for several decades. However, after decades of research we do not have a definite answer on what the driving factors are of such peripheral choices, but rather mixed evidence based on different economic and political factors. The seminal analysis of \citet{eichengreen2019mars}, conducted in a set of countries before World War I, has provided a dual theoretical framework of economic and security/geopolitical factors that are having an effect on the currency composition of foreign exchange reserves. In this study, scholars employ two main approaches to analyze which currencies should be used in international transactions and as savings in deposits \citet{eichengreen2019mars}. In the so-called Mercury hypothesis, scholars focus on economic costs and benefits, safety, liquidity, inflation forecasts, as in \citet{romer2000federal}, when explaining why some currencies are used more than others in cross-border transactions, e.g. \citet{eichengreen2008rise}. On the other side of the argument,  scholars also hypothesize the so-called Mars hypothesis, which is focused on geopolitical elements such as the role of strategic and military-power have on others when governments choose reserve currencies. Their estimates are based on the hypothesis that military power provides a stabile economic environment and less exchange rate volatility of that particular economy, so there is a great deal of interconnectedness of political and economic factors. In addition, the leading countries can exert successful political and military hegemony based on which they encourage other countries to use their currencies \citet{eichengreen1985exchange}. The literature has existent evidence on the economic factors affecting the decision to hold foreign reserves in a particular currency, but it is scarce on evidence of geopolitical factors, such as political influence and military strength, shaping decisions on foreign exchange reserves. The main reason is the lack of causal identification and the complexity of defining geopolitical factors such as political or military power through a single variable or within a model.
In this study, we provide a network theory model to show under different parameter values how different political factors, such as the same geo-political allegiance may have even stronger effect on choosing the dominant foreign exchange currency than the economic factors. As \citet{eichengreen2019mars} stated, the Mars hypothesis may vividly explain why Japan, Germany, the Netherlands, Belgium have a higher percentage of their foreign reserves in dollars in contrast to France, for instance, a country with nuclear weapons. It seems that a stronger military power can diversify its portfolio in its foreign reserves much more than just a common non-nuclear US ally dependent on the United States for security. With the turbulent geopolitical tensions between the US and the EU around NATO involvement, this topic has become more important than ever. Not only for NATO members, but also for the entire world. The Mars hypothesis of geopolitical factors being more important than economic may be further supported by the fact that during the last few years China has been rapidly disposing of the dollars in its foreign reserves, as shown in Figure \ref{CHNUS}. Similarly, Russia has been reducing the amount of holdings in dollars since starting their initial invasion in Ukraine in 2014. As shown in Figure \ref{RUSUS} it has been steadily dropping from year to year. In contrast, even though India created BRICS together with Russia and China, its attitude towards holdings of US Treasury's is completely different, it has been increasing over the past decade (see  Figure \ref{INDUS}). Looking at only these three geopolitical superpowers in relations to the US we cannot disentangle whether economic factors were the main reason of the change in holdings of US Treasuries or geopolitical factors, such as invasion of Ukraine, change in the US rhetoric towards China, or something else. When politicians in country A start openly saying that country B is a threat and even an enemy, savings and trading are no longer just an economic issue, but rather a broader political issue. In that case a larger weight is put on the long run geopolitical goals, than on short term economic goals. US treasuries may be a much safer investment, but countries will not support strengthening of economic power of their foes, on the contrary, they will do everything in their power to weaken them economically. 
    
There is no place on Earth where tigers and lions live together because they are predators of the same level. If by chance they met as random walkers, one would tear the other apart, even though they had never met before. Indeed,  until a couple of years ago even the largest US opponents, Russia and China, when they trade through the West-controlled SWIFT payment system \citet{eichengreen2024international}, the payments were realized mainly in dollars, ultimately helping the U.S. economy and particularly making its currency stronger.  In a homogeneous world where there is only a single dominant power, even the largest competitors partially pay the bill. Generally, the more countries save and trade in a country's  currency,   the stronger the currency, but also weaken the inflation shocks once the country prints new money during recessions, lockdowns, and wars. Therefore,  the dominance of the world leader is maintained not only through the country's military power but also through controlled financial and banking systems.

In this study, we investigate a complex interaction between country`s military spending, GDP, geopolitical allegiance and consequently share of foreign exchange reserves kept in their allies currency. Similarly as in the rest of the literature, we acknowledge that the direction of causality is hard to disentangle without a proper identification strategy. For that reason, the first part of our empirical analysis is used to show the general direction and correlations between the main variables. The novel value-added part is in the second part of our analysis, within our network theory model. We introduce a novel geopolitical shock into a model which is not economically driven but purely political, like the change of a geopolitical block and how it affects long-run network dynamics. 

In the first part, we analyze the linear relationship between the currency composition of Official Foreign Exchange Reserves (COFER) collected by the International Monetary Fund (IMF) and GDP and MS of some western-most countries, China, and a group of other aggregated countries for a time period ranging from 2003 to 2023. We observe a rather stable regression relationship between (the logarithm of) Foreign Exchange Reserve Composition and both (the logarithm of) GDP and (the logarithm of) MS of western countries. Data on the Chinese renminbi (RMB) and on aggregated other currencies with the corresponding GDP or military spending are outliers with respect to the linear relationship between FER versus GDP and FER versus MS. The yearly analysis performed during the period from 2016 to 2023 (that is, since when the IMF report data for Chinese RMB) shows that the role of Chinese RMB in (a) FER and GDP and (b) FER and MS of China progressively converges to the level of linear relationship typically observed for western countries. To investigate the role of multicollinearity in regression analysis, we apply methods such as Ridge and Lasso as in \cite{james2013introduction} and verify that MS is more explicative than the GDP level for the seven western countries (Australia, Canada, EU, Japan, Swiss, UK, US). Similarly as in the research so far, we acknowledge all the limitations our study may have towards a more causal interpretation of our estimates.    
Inspired by the Mars hypothesis, we report that a country's military spending, taken as a proxy for the size of a country, highly correlates with the country's currency in Foreign Exchange Reserves (FER), where the Chinese RMB is well outside the FER equilibrium, but there is a steady trend towards a new FER balance.  If this gradual and not abrupt trend continues in the coming years, the RMB may reach the new FER balance within approximately 15 years. In the model, once China’s GDP overtakes that of the U.S., the U.S. responds by implementing sanctions and imposing restrictions on its domestic market. In turn, China begins a process of dedollarization, which aligns well with empirical data.

Our empirical findings are supported by our findings within a network theory model.


\section{Literature review} 

The accumulation of foreign exchange reserves (FER) has been a prominent feature of economic policy, especially among emerging market economies (EMEs), over the past several decades. The reasons for the accumulation of FER can be both political and economical. Our paper is on the intersection of political and economical motives, so we will provide insights into both strands of literature, as well as theoretical and empirical evidence. From the early 1990s to 2018, according to \citet{arslan2019size} emerging market economies increased their FX reserves from approximately 5\% to nearly 30\% of GDP. Historically, the US dollar has been a prominent currency for FER, but in the last decade we have seen significant shifts from that equilibrium. The shift in some countries is motivated by economic factors, while others are motivated purely geopolitically, as shown in Figures \ref{CHNUS}, \ref{RUSUS}, and \ref{INDUS}.

Different authors provide different economic reasons for the use of FER in a particular currency, and these reasons were evolving differently over time and geographical regions. In terms of economic motives, several factors have been involved, including precautionary motives, export competitiveness, and the pursuit of macroeconomic stability. The holding of substantial FER reserves can serve as a buffer against external shocks, reducing the volatility of the exchange rate, and fostering a stable economic environment conducive to growth. It is important to note that the findings in the literature are mixed and highly dependent on the context: the type of shock, the countries involved in a study, the time period and the methodology used to infer the main findings. 

\citet{ito2020currency} analyses the factors that govern the choice of the currency composition of official foreign exchange reserves. They provide evidence that the currency composition of reserves is strongly related to the comovement of the domestic currency with key currencies and the currency invoice of trade. This is an important feature, which we also build in our theoretical model, since "trade wars" are one type of shock affecting the probability to switch to a different currency in FER. 

FER have also been used as a tool to maintain macroeconomic stability and have been supported by empirical work in the literature. \citet{ahmed2023effectiveness} show that an additional 10 percentage points of FX reserves in GDP are associated with a reduction of 1.5\% to 2\% in currency depreciation during periods of global financial stress. Similarly,
\citet{dominguez2012international} find that higher reserve accumulations before the crisis are associated with higher post-crisis GDP growth. More specifically, a contrary view \citet{alwadeai2024beyond} investigates the impact of economic sanctions on the volatility of exchange rates, with a specific focus on the role of foreign reserves in mitigating these effects. They find that high reserves-to-GDP ratios do not fully stabilize exchange rate volatility in the presence of economic sanctions, challenging the traditional view of reserves as reliable stabilizers.

However, according to \citet{rodrik2006social}  the benefits of maintaining large FER reserves must be balanced against the associated costs. The opportunity cost of holding reserves is significant, as these funds could otherwise be invested in domestic projects to stimulate economic growth. Moreover, excessive accumulation of reserves may lead to issues such as inflationary pressures and asset bubbles. Similarly, \citet{levy2020cost}  illustrate that the cost of holding reserves may have been considerably lower than is generally assumed in both the academic literature and the policy debate. In order to establish a benchmark, \citet{jeanne2011optimal} provide a framework for determining optimal reserve levels for emerging market economies based on various economic parameters. A benchmark calibration suggests that an optimal level of reserves for EMEs is approximately 10.1\% of GDP, or 92\% of short-term external debt, which is close to the average ratio observed between 1975 and 2003.

Unlike most economics models which consider only an economic framework to calibrate the model, our theoretical model considers geopolitical shocks as both of economic or purely political nature. For example, Russia`s invasion of Ukraine has been a military move that has had an enormous effect on cutting off geopolitical ties. For that reason, in our model we introduce different variables of military strength, which, in addition to economic strength, also have a crucial effect of positioning in the global FER market. The interconnection between FER reserves, GDP, and military spending is complex and multifaceted. Economic strength, as reflected in GDP, provides the foundation for both accumulating FER reserves and sustaining military expenditures. 

The first strand of the literature connects different economic measures, mainly GDP, with military spending. \citet{clements2021military} find that military spending in relation to GDP does not reach a common level in 138 countries in their sample. It is divided into three main groups. The first group of 20 countries experiencing a high degree of conflict, spending has increased and diverged from the global trend. The second group has the largest number of countries - 77, of which 30 are advanced economies - generating 90 percent of global military spending. Their average military spending in relation to GDP fell significantly from 1990 through the mid-2000s, but has changed little since then. This group includes China, India, Russia, the United Kingdom and the United States. Furthermore, \citet{barnum2025measuring} find two interesting findings related to our research. First, correlations between economic surplus and military spending are currently higher than at any point in the last 200 years. Second, find that the (negative) effect of a democratic ally on military spending is three times larger, and the (positive) effect of an increase in GDP is five times larger than previously estimated. 

Military expenditure is a critical component of a nation's defense strategy, directly influencing its military capabilities and power projection. The relationship between military spending and economic growth has been extensively studied, yielding mixed findings. Some research suggests that increased defense spending can stimulate economic activity through job creation and technological advancements. However, other studies argue that excessive military expenditure can crowd out public investment in other sectors, potentially hindering economic growth. For example,
\citet{peltier2023we} provide an in-depth analysis on how decades of high levels of military spending have changed the US government and society, strengthening its ability to fight wars while weakening its capacities to perform other core functions. They highlight in their study how investments in infrastructure, healthcare, education, and emergency preparedness, for instance, have all suffered as military spending and industry have crowded them out.

The effectiveness of military spending in enhancing military power also depends on factors such as strategic allocation of resources, efficiency of defense institutions, and the existing security environment. The mere increase in the defense budget does not automatically translate into greater military effectiveness; the quality of expenditure and the geopolitical context play a pivotal role. In addition, political accountability and transparent governance are essential in ensuring that military spending contributes positively to national security without undermining economic stability.

Historical analyses, such as Paul Kennedy's "The Rise and Fall of the Great Powers," illustrate that nations with disproportionate military expenditures relative to their economic base often experience "imperial overstretch," where the burden of defense commitments exceeds the country's economic capacity, leading to decline. This underscores the importance of aligning military ambitions with economic realities to ensure sustainable national power. Moreover, \citet{modelski1996leading} argue that naval power is the key tool that allows world political leadership, colonization, and domination in world trade. However, a less explored topic in most studies of hegemonic patterns is the military expenditure factor in the competition between states for military and economic dominance. The leader nation increases its military capabilities to maintain its dominant position, while the alliance states support the leader in order to benefit from greater investments in their economies, as in \citet{modelski1996leading}. Regarding the link between military and economy, \citet{kennedy1989rise} pointed out a significant correlation between revenue raising capacities and the military strength of the country. Paul Kennedy also argued that military spending by hegemonic countries can become a burden on its economy, eventually heading to economic collapse. This argument seems to be controversial due to the lack of economic analyses that would support his basic idea of the interaction between military spending and economic growth. 

The complexity of the interaction between military and economic power and the subsequent connection to the use of a particular global power as the main currency lacks serious causal connections in the research. History has been full of changes in military and economic power, yet how these changes affected changes in foreign exchange reserves is relatively understudied. For many decades during which the United Kingdom had been dominating the entire world, the UK sterling was the leading currency, just to be later changed by the US dollar. For example, \citet{triffin1960gold} estimated that not only in 1928, the share of pound sterling of global foreign exchange reserves was around 80 percent, but  even in 1938  pound starlings share remained around 70 percent. More recently, \citet{chinn2008euro}, suggested that dollars only overtook sterling after World War II. Based on data gained from central bank archives, \citet{eichengreen1985exchange} and \citet{eichengreen1986competitive}, reported that the dollar overtook sterling as the leading global reserve currency in the mid-1920s, not in 1928, 1938 or 1948.

One of the first historical investigations to show the connection between foreign currency reserves and military alliances was \citet{eichengreen2019mars}. Applying data on foreign currency reserves of 19 countries before World War I, for which the currency composition of reserves and military and security alliance memberships are known, \citet{eichengreen2019mars} validated the hypotheses of Mercury and Mars. They demonstrated that military alliances increase the currency percentage in the ally's foreign reserve assets by 30 percentage points. \citet{eichengreen2019mars} argued about why a country decides to use an ally's currency by linking the national security and military alliance on one side and foreign policy planning of foreign currency reserves on the other. Reserve currency issuers such as the US may use military-dependent nations to obtain finance from them by issuing treasuries or, as in the case of the US, the nuclear power as the financial center may borrow funds to their allies. 

Beyond solely geopolitical alliances, there are other factors to keep an ally s currency. In articles on foreign reserve currency, financial depth and development is an important factor \citet{portes1998euro}. Moreover, \citet{chitu2014dollar}  demonstrate that financial development was the most important factor in helping the dollar overcome sterling's dominance as an international financing currency during the interwar period.  The credibility of the reserve currency issuers is motivated by theoretical models in which agents in non-dollar economies trade indirectly using the US dollar instead of using direct bilateral trade among their own currencies if these currencies depreciate too often and for too long and there are substantial transaction costs of exchange according to \citet{devereux2013vehicle}. Economic size is another standard factor of international currency choice. \citet{krugman1980vehicle} discussed the effects of the network, focusing on the increased returns resulting from economies of scale. The economy of scale was a topic in the random matching model of \citet{matsuyama1993toward}, who model the choice of international currencies in a way that the incentive of an agent to accept a country's currency depends on the frequency of trades with a given country. Furthermore, the strategic use of FER reserves can influence a nation's defense posture. For example, countries with substantial reserves may have greater flexibility in responding to economic sanctions or funding defense initiatives without resorting to external borrowing. However, the opportunity cost of holding large reserves, as opposed to investing in economic development, must be carefully considered. Thus, policymakers must navigate the delicate balance between maintaining sufficient FER reserves, fostering economic growth, and allocating resources to defense to ensure comprehensive national security.

In summary, the literature indicates that while FER reserves and military spending are tools for achieving economic stability and national security, their optimal levels and effectiveness are contingent upon a nation's economic capacity and strategic priorities. Policymakers must consider the intricate interplay between these elements to formulate policies that promote sustainable growth and robust defense capabilities.


\section{Data.}~In the paper we  use Allocated Reserves of Forex exchange reserves, where in a strict sense, FER includes only deposits of foreign currencies held by nationals and monetary authorities. The data are provided by the IMF, specifically claims in US dollars, Euro, Chinese RMB, Japanese yen, British pound sterling, Australian dollars, Canadian dollars, Swiss francs, and aggregated claims in other currencies. The data are listed quarterly, and in the paper we always use the last Q4 quarter. Military spending data are provided by the World Bank. The GDP data recorded annually are provided by \citet{imfdata2024}.


\section{Methodology}

\subsection{Empirical strategy}

\citet{eichengreen2019mars} used data before World War I that analyze the composition of the currency of foreign exchange holdings by countries, focusing on the pairwise interaction between the currency reserve issuer and the holder of the reserve currency. We, on the contrary, used the most recent decade (2013 to 2023) to analyze the aggregate foreign exchange reserves and varying economic variables of a set of developed countries which are the reserve issuers, without asking which countries hold the money. 

For a set of Western currencies, we test the hypothesis that the Military Spending (MS) taken as a proxy for the size of a country strongly correlates with the Foreign Exchange Reserves (FER). To this end, precisely, for the dollar, euro, British sterling, Japanese yen, Swiss franc, Australian and Canadian dollars, for which the IMF provides data on Foreign Exchange Reserves, \citet{imfdata2024},  we apply Simple Regression  Analyses  (see Fig.  \ref{regressionMSFER}). 
For each year $t$ we apply  simple regression  for a set of countries $i$
   \begin{equation}
ln(FER)_{i,t} = \beta_0   +    \beta_1     ln( MS) _{i,t} + e_{t}, 
\label{e2}
\end{equation}   
where, $MS$ represents  Military Spending and   $e_i$ is the random error. The MS variable can be considered as a proxy for a size of a country, but also a proxy for alliance variable, since all countries analyzed at this stage are the Western U.S. allies.

However, GDP and MS are highly correlated, and next we apply methods such as Ridge and Lasso to take into account multicollinearity in multiple regression data as in \citet{james2013introduction}. 
 
Ridge method defined as
\begin{equation}
  \beta_{\lambda} =     arg  ~ min  \Sigma_ {i=1} ^{N}   (y_i  - x_i b) ^{2}   +\lambda~ \Sigma_ {i=1} ^{k} b_ {i} ^{2} , 
\label{e2}
\end{equation}     
is most suitable when a data set contains a higher number of predictor variables than the number of observations. In the above formula $b$ are coefficients, $k$ is the number of coefficients and $N$ stands for the number of data points, where the penalty parameter $\lambda $ is a positive constant. Note that the variance of the ridge estimator is always smaller than the variance of the OLS estimator. Since the ridge estimator is not scale invariant, in analyses we must standardize all the variables in regressions by subtracting from each variable its mean and dividing it by its standard deviation. Both Ridge and Lasso regressions reduce the coefficient estimates toward zero. However, Lasso  regression defined as 
\begin{equation}
  \beta_{\lambda} =     arg  ~ min  \Sigma_ {i=1} ^{N}   (y_i  - x_i b) ^{2}   
  +\lambda~ \Sigma_ {i=1} ^{k} | b_ {i}| , 
\label{e3}
\end{equation}    
reduces some of the coefficients  exactly  to zero when the tuning parameter becomes large enough.  Thus, in general, Ridge regression is used when the goal is to minimize the impact of less important features while keeping all variables in the model. In contrast, 
Lasso regression is preferred when the goal is feature selection with fewer variables, making a more interpretable model.

Motivated by the literature on foreign currency reserves \citet{li2008rmb} and \citet{eichengreen2019mars}, we  model  the  reserve currency as:
\begin{multline}
ln(FER)_{i,t} = \beta_0   + 
 \beta_1  ln( MS) _{i,t} +     \beta_2'   \textbf{X}_i        
+  \beta_3   DV ^{RMB}_{i,t}  +   \beta_4    DV^{Other}_{i,t}   +  e_{i,t}, 
\label{e5}
\end{multline}   
where  $i$  represents   the   country  index and $t$ is  time dimension $( i = 1...9;  t = 1...20)$,  
 $X$ is a vector of control variables as varying as 
 gold reserves,   international trade and financial depth; 
 $MS$ is Military Spending of country $i$; 
 DVs are  two  dummy variables   denoted to RMB and Other currencies  equivalent to alliance variable 
 in \citet{eichengreen2019mars}; 
 $e$ is the residual and the $\beta$s are the regression's coefficients.

Economic size is an additional standard determinant of the choice of international currencies. Empirically, we measure economic size by the Military Spending that is proportional to the GDP and, therefore, the output of the country issuing reserve currency.

Among control variables, financial depth is an important factor in the choice of reserve currency \citet{portes1998euro}. Motivated by \citet{king1993finance}, we quantify the financial depth by the financial monetization ratio (wide money M2 to GDP), taking data from the World Bank \url{(https://data.worldbank.org/indicator/FM.LBL.BMNY.GD.ZS)}.

We implement Equation (\ref{e5}) using multiple regression and report the values of $p$. If geopolitical considerations dominate,  we expect the coefficient on economic size to be statistically significant. Hence, our test for supporting the '' Mars''  hypothesis  is:
\begin{equation}   
H_0:  \beta_1  >  0,  \beta_2  >  0,
\label{e6}
\end{equation}   
where accepting $H_0$  is evidence for the Mars hypothesis.

\section{Results.}

 In this first subsection, we provide the results of our empirical analysis. In the second part, we provide the results of our network theory model.

\subsection{Empirical}

\begin{figure}[!hb]
  \centering 
   \caption{Regression  between Military Spending and Foreign Exchange Reserves for 2022. Regressions 
   obtained in ln-ln plot in billions of dollars.  }\includegraphics[width=0.9\textwidth]{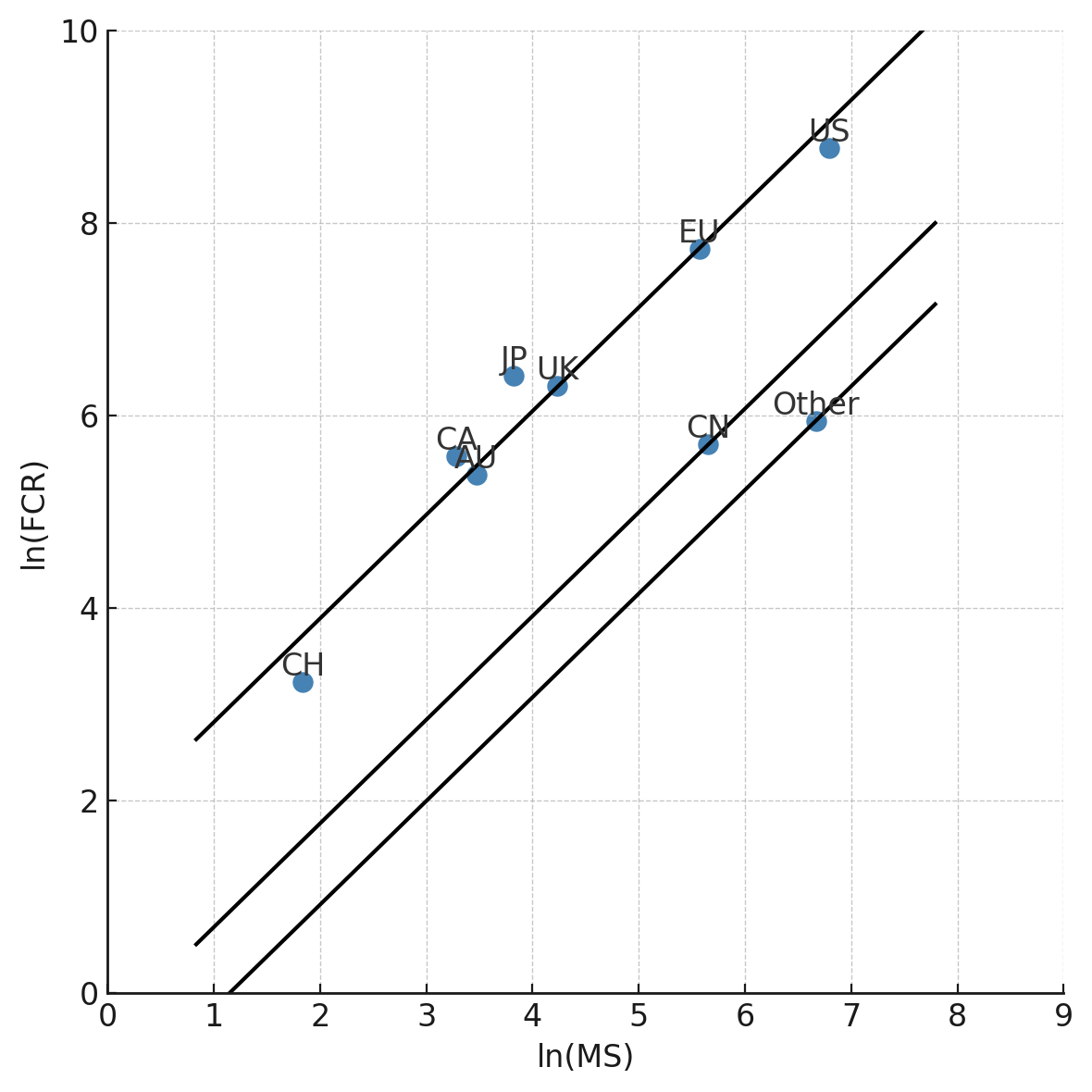}\\
 
  \label{regressionMSFER}
\end{figure}

For equation 1 we report the regression coefficients for each year separately between 1993 and 2023, the country's currency enlisted in Foreign Exchange Reserves highly
significantly depends on the country's Military Spending (MS)  with
surprisingly high  $R^2 $  from 0.96 to 0.99, for each year (Table 1, last column). Surprisingly high $R^2 $  may support the Mars hypothesis, suggesting that western FER currencies have already reached an FER equilibrium between the average MS and the fraction of FER.  Note that all currencies in our analysis belong to US allies and, therefore, the set represents a proxy for a variable alliance, as suggested by Eichengreen, Mehl, and Chitu\cite{eichengreen2019mars}. However, here we note that the number of currencies listed by FER has been changing over the years, as the Canadian and Australian dollars appeared in 2012.

\begin{table}[!htbp]
\begin{center}
\caption{Regression coefficients (standard errors) from log-log models estimating the relationship between Foreign Exchange Reserves (FER) and Gross Domestic Product (GDP), with and without the inclusion of the shares of Chinese RMB and Other Currencies in total reserves.}
\label{t2}
\begin{tabular}{c@{\hskip 12pt}c@{\hskip 12pt}c@{\hskip 12pt}c@{\hskip 12pt}c@{\hskip 12pt}c}
\hline
$ $ & $\beta_0$ & GDP & China & Others & $R^2$ \\
\hline
2023 & -5.425 (0.0073) & 1.343 (---) & --- & --- & 0.933 \\
2022 & -5.283 (0.0118) & 1.367 (0.00037) & -2.569 (0.00698) & -4.719 (0.00146) & 0.935 \\
2021 & -5.616 (0.01868) & 1.428 (0.00076) & -2.547 (0.01509) & -4.876 (0.00281) & 0.914 \\
2020 & -5.368 (0.02188) & 1.416 (0.00083) & -2.617 (0.0143) & -4.966 (0.00278) & 0.911 \\
2019 & -5.957 (0.01350) & 1.468 (0.00062) & -2.726 (0.0107) & -5.135 (0.00212) & 0.921 \\
2018 & -5.284 (0.01254) & 1.500 (0.00066) & -2.711 (0.01207) & -5.201 (0.00228) & 0.919 \\
2017 & -5.708 (0.01622) & 1.438 (0.00072) & -3.020 (0.0069) & -5.061 (0.00230) & 0.916 \\
2016 & -5.945 (0.02052) & 1.445 (0.00116) & -3.020 (0.00116) & -5.021 (0.0038) & 0.898 \\
2015 & -5.809 (0.00755) & 1.344 (0.00034) & --- & --- & 0.937 \\
2014 & -6.527 (0.00513) & 1.376 (0.00030) & --- & --- & 0.940 \\
2013 & -6.429 (0.00484) & 1.336 (0.00028) & --- & --- & 0.942 \\
\hline
\end{tabular}
\footnotesize
\item \textit{Notes}: Values are reported as coefficient (standard error). "---" indicates missing values. $\beta_0$ is the intercept. GDP refers to U.S. Gross Domestic Product. "China" and "Others" represent currency shares in global reserves. $R^2$ indicates model fit. All values are expressed in billions of U.S. dollars.
\end{center}
\end{table}

MS of a given country is somewhat proportional to its GDP, and therefore it is reasonable to assume that the fraction of the country's GDP in the world's GDP mainly determines how popular the country currency is in foreign exchange reserves. Indeed, the fractions of pounds, yen, euro, and both Australian and Canadian dollars in Foreign Exchange Reserves are just a bit larger than are the fraction of country's GDP in the world's GDP. Only in the case of the dollar, that currency is substantially more represented in the Reserves than is the US fraction in the World GDP. In contrast, the Chinese RMB is far from the balance, but with a steady increase towards the potential future balance. Next, we report that not only Military Spending, but also GDP (see Table \ref{t2}) is in strong correlation with foreign exchange reserves. Comparing $R^{2}$ in Table \ref{t2} and Table \ref{t3}  we report that MS is also more influential than the overall country fraction of international trade. Recalling that the variables are standardized, Table \ref{t4} reports that MS is more influential than GDP for each year considered.

\begin{table}[!htbp]
\begin{center}
\caption{Regression coefficients (standard errors) from log-log models estimating the relationship between Foreign Exchange Reserves (FER) and international trade (exports and imports), including the shares of Chinese RMB and Other Currencies.}
\label{t3}
\begin{tabular}{c@{\hskip 12pt}c@{\hskip 12pt}c@{\hskip 12pt}c@{\hskip 12pt}c@{\hskip 12pt}c}
\hline
$ $ & $\beta_0$ & Trade & China & Others & $R^2$ \\
\hline
2023 & -3.204 (0.399) & 1.251 (0.041) & -1.200 (0.414) & -3.106 (0.130) & 0.608 \\
2022 & -3.598 (0.313) & 1.291 (0.027) & -1.995 (0.189) & -3.155 (0.096) & 0.659 \\
2021 & -3.518 (0.355) & 1.308 (0.036) & -2.051 (0.220) & -3.228 (0.114) & 0.622 \\
2020 & -3.338 (0.372) & 1.316 (0.037) & -2.171 (0.199) & -3.398 (0.103) & 0.624 \\
2019 & -4.025 (0.288) & 1.376 (0.029) & -2.222 (0.182) & -3.466 (0.093) & 0.655 \\
2018 & -4.203 (0.269) & 1.390 (0.028) & -2.210 (0.186) & -3.502 (0.092) & 0.659 \\
2017 & -3.783 (0.300) & 1.344 (0.029) & -2.585 (0.121) & -3.425 (0.089) & 0.667 \\
2016 & -3.844 (0.310) & 1.334 (0.036) & -2.604 (0.134) & -3.319 (0.108) & 0.643 \\
\hline
\end{tabular}
\footnotesize
\item \textit{Notes}: Values are reported as coefficient (standard error). Trade refers to the sum of exports and imports in billions of U.S. dollars. "China" and "Others" represent the respective currency shares in global reserves. $R^2$ indicates the model's explanatory power. Although the models show relatively high $R^2$, they are notably lower than models using military spending as a predictor. All values are expressed in billions of U.S. dollars.
\end{center}
\end{table}

Furthermore, we report that the Chinese RMB brings a distortion in the financial FER market where over time the entire market aims to a new equilibrium. Thus, next, we study how the Chinese currency has been changing over the years with the goal of predicting when the RMB may potentially reach a new FER equilibrium. Currently, the Chinese currency can be considered outside the FER equilibrium because the Chinese GDP in nominal dollars is around 17\%, while the RMB contributes less than 3\% in the foreign exchange reserves.

Next, for each year between 2016 and 2022 for which the data for RMB are available,  we add the Chinese RMB by introducing a dummy variable (DV) where its value is 1 in case the currency is RMB and 0 otherwise. Similarly, we also introduce a dummy variable for all other FER currencies, where again its value is 1 in case the currency is Others and 0 otherwise. For each year applying multiple regression for a set of countries $i$
\begin{equation}
ln(FER)_{i,t} = \beta_0  +   \beta_1  ln( MS) _{i,t} +    \beta_2   
DV ^{RMB}_{i,t}  +    \beta_3       DV^{Other}_{i,t}  +  e_{i}, 
\label{e2}
\end{equation}   
where, $MS$ represents  Military Spending,  where two DVs  are denoted to RMB and Other currencies aggregated  on one, and   $e_i$ is the random error. First, note that two dummy variables do not change the regression coefficients obtained for simple regressions (see Fig. \ref{regressionMSFER})

\begin{figure}[!hb]
  \centering 
   \caption{Deposit of gold in a set of developed countries.}
   \includegraphics[width=0.9\textwidth]{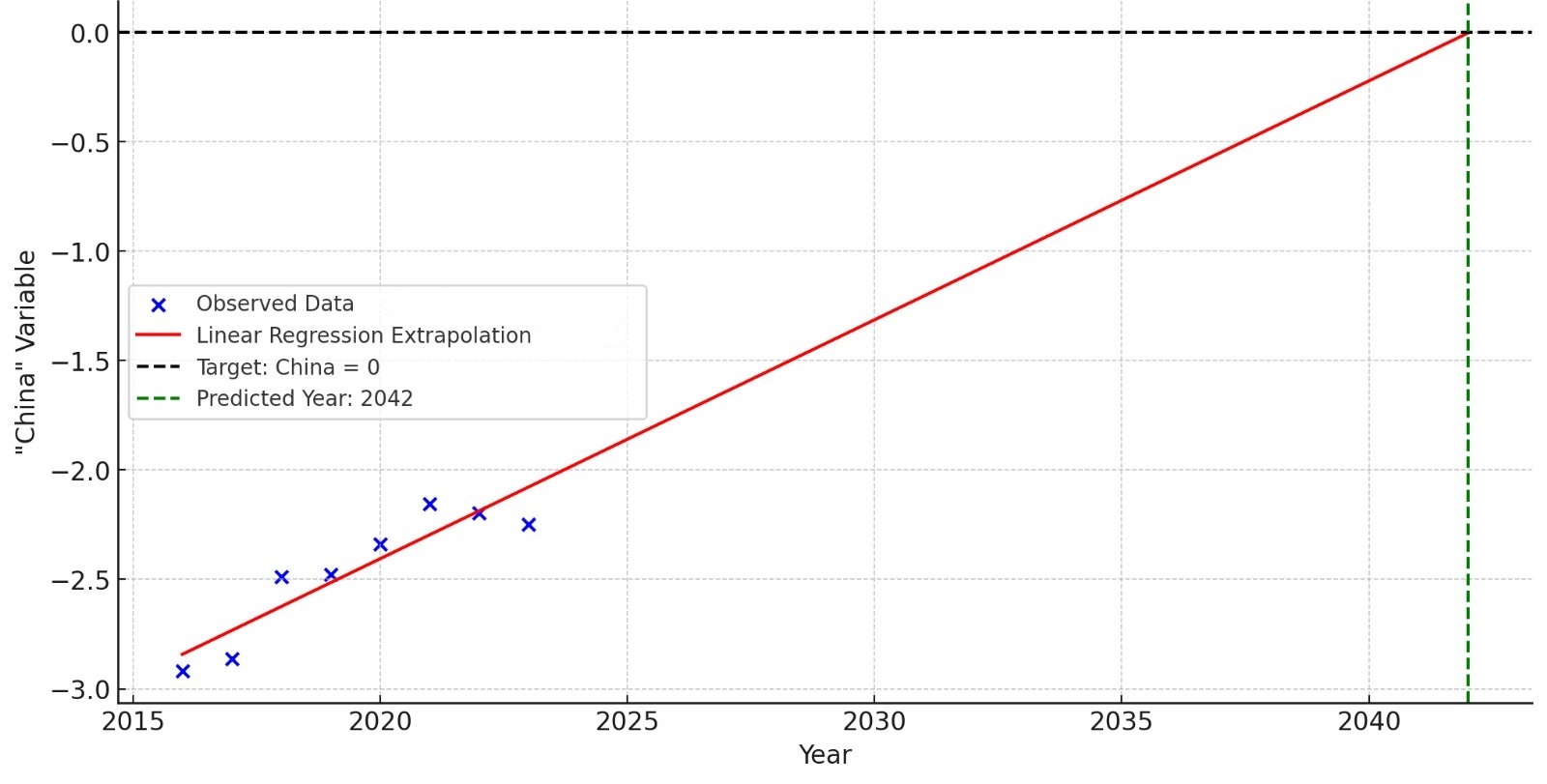}\\
 
  \label{extrapol}
\end{figure}

In Table 1, we see that the coefficient of the dummy variable is negative as one may expect because compared with the Western counterpart currencies, the RMB is underrepresented in FER relative to the country's GDP. Second, consistently over the years the Chinese RMB coefficient has been ramping down in absolute terms bringing the FER market closer to the new FER equilibrium where roughly the share of the country's currency is in good agreement with the country's MS and roughly the GDP. Extrapolating the data on the RMB dummy coefficient in Table 1, roughly, one can predict that the RMB could reach an equilibrium with the currencies of western countries where it will be considered a regular FER currency in $ \approx  $ 18 years (see Fig. \ref{extrapol}). The prediction assumes that in years to come the world will experience gradual changes in relationship between the U.S. and China, and not the breakdown-in-relations scenario in which the world financial system is heading to a tipping point. In contrast to the RMB dummy coefficient, the dummy coefficient corresponding to Others exhibits stability to a great extent in the same years. It is important to note that this is based on linear trend and is a simplification because year-to-year ceteris paribus conditions of other variables will have a differential effect on the trend.

At the 0.05 level, for all years, the regression coefficients are jointly significant (F-test), and the slope and dummies coefficients are individually significant (t-test). Over time the regression coefficients exhibit stability, where only during the last two years characterized by the Russian invasion the regression coefficients slightly ramped down. However, the intercept coefficients are individually significant only after 2012 (the IMF data exclude AUD, CAD before 2012). Furthermore, at 0.05 level, for all years, homoskedasticity is not
rejected (Breusch-Pagan), while the stability (equality) of coefficients between the groups is rejected (Chow).

\begin{table}[!htbp]
\begin{center}
\caption{Regression coefficients from Ridge and Lasso models estimating the relationship between Foreign Exchange Reserves (FER), Military Spending, and GDP for seven Western currencies.}
\label{t4}
\begin{tabular}{cc@{\hskip 12pt}c@{\hskip 12pt}c@{\hskip 12pt}c}
\hline
$ $ & Model & Military Spending & GDP & $R^2$ \\
\hline
2022 & Ridge & 1.232 & 0.389 & 0.965 \\
     & Lasso & 1.236 & 0.384 & 0.965 \\
\hline
2021 & Ridge & 1.433 & 0.261 & 0.959 \\
     & Lasso & 1.438 & 0.256 & 0.959 \\
\hline
2020 & Ridge & 1.440 & 0.264 & 0.957 \\
     & Lasso & 1.443 & 0.259 & 0.957 \\
\hline
2019 & Ridge & 1.548 & 0.221 & 0.972 \\
     & Lasso & 1.552 & 0.216 & 0.971 \\
\hline
2018 & Ridge & 1.627 & 0.171 & 0.978 \\
     & Lasso & 1.632 & 0.165 & 0.978 \\
\hline
2017 & Ridge & 1.570 & 0.158 & 0.983 \\
     & Lasso & 1.574 & 0.153 & 0.983 \\
\hline
2016 & Ridge & 1.706 & 0.040 & 0.980 \\
     & Lasso & 1.780 & 0.036 & 0.981 \\
\hline
2015 & Ridge & 1.388 & 0.262 & 0.995 \\
     & Lasso & 1.391 & 0.258 & 0.995 \\
\hline
2014 & Ridge & 1.294 & 0.401 & 0.999 \\
     & Lasso & 1.296 & 0.398 & 0.988 \\
\hline
2013 & Ridge & 1.238 & 0.425 & 0.991 \\
     & Lasso & 1.240 & 0.422 & 0.981 \\
\hline
\end{tabular}
\end{center}
\caption*{\footnotesize\textit{Notes}: Military Spending refers to global annual defense expenditures in billions of U.S. dollars. GDP represents Gross Domestic Product in billions of U.S. dollars. $R^2$ denotes the coefficient of determination, indicating model fit. Ridge and Lasso refer to regularized regression techniques applied to improve model generalization.}
\end{table}

\begin{table}[!htbp]
\begin{center}
\caption{Regression coefficients (standard errors) from log-log models estimating the relationship between Foreign Exchange Reserves (FER) and key macroeconomic variables.}
\label{t6}
\setlength{\tabcolsep}{6pt}
\begin{tabular}{cccccccc}
\hline
Year & $\beta_0$ & MS & Gold & IT & M2 & CHN & $\tilde{R}^2$ \\
\hline
2022 & -3.04 (0.02) & 0.70 (0.00) & -0.18 (0.01) & 0.18 (0.06) & 0.66 (0.01) & -2.84 (0.00) & 1.00 \\
     & 0.60 (0.76) & 1.07 (0.01) & -0.13 (0.34) & 0.22 (0.56) & --- & -2.21 (0.02) & 0.99 \\
     & 1.73 (0.02) & 1.15 (0.00) & -0.08 (0.38) & --- & --- & -2.14 (0.01) & 0.95 \\
2021 & -3.09 (0.05) & 0.85 (0.01) & -0.24 (0.02) & 0.23 (0.13) & 0.57 (0.03) & 2.76 (0.00) & 1.00 \\
2020 & -2.74 (0.19) & 0.91 (0.02) & -0.24 (0.07) & 0.22 (0.36) & 0.52 (0.09) & -2.86 (0.00) & 0.99 \\
2019 & -2.77 (0.11) & 0.92 (0.01) & -0.22 (0.04) & 0.23 (0.21) & 0.49 (0.05) & -2.96 (0.00) & 0.99 \\
     & -0.03 (0.99) & 1.18 (0.00) & -0.17 (0.23) & 0.26 (0.49) & --- & -2.53 (0.01) & --- \\
     & 1.38 (0.02) & 1.27 (0.00) & -0.11 (0.24) & --- & --- & -2.49 (0.00) & --- \\
2018 & -2.62 (0.05) & 0.98 (0.00) & -0.21 (0.02) & 0.25 (0.09) & 0.43 (0.03) & -2.93 (0.00) & 1.00 \\
2017 & -1.79 (0.05) & 0.99 (0.00) & -0.18 (0.01) & 0.23 (0.05) & 0.34 (0.02) & -3.21 (0.00) & 1.00 \\
2016 & -1.53 (0.09) & 1.10 (0.00) & -0.22 (0.01) & 0.35 (0.03) & 0.14 (0.15) & -3.12 (0.00) & 1.00 \\
     & -0.81 (0.25) & 1.17 (0.00) & -0.21 (0.01) & 0.37 (0.04) & --- & -3.00 (0.00) & 1.00 \\
     & -1.20 (0.01) & 1.28 (0.00) & -0.12 (0.10) & --- & --- & -2.95 (0.00) & 0.98 \\
\hline
\end{tabular}
\footnotesize
\item \textit{Notes}: MS = Military Spending; Gold = Gold Reserves; IT = International Trade (exports + imports); M2 = Broad money supply; CHN = Share of Chinese RMB in global reserves. All variables are expressed in logarithmic form and in billions of U.S. dollars. Coefficients are shown with standard errors in parentheses. $\tilde{R}^2$ indicates model fit. Rows without a year are alternate specifications for the same year.
\end{center}
\end{table}

\begin{table}[!htbp]
\begin{center}
\caption{Results of Breusch–Pagan (LM) and F-tests for heteroskedasticity in residuals from FER regression models, 2016–2022.}
\label{t7}
\begin{tabular}{ccccc}
\hline
Year & LM Statistic & LM $p$-value & F-statistic & F $p$-value \\
\hline
2022 & 7.813 & 0.166 & 16.73 & 0.057 \\
2021 & 4.325 & 0.503 & 0.471 & 0.785 \\
2020 & 5.632 & 0.344 & 0.952 & 0.583 \\
2019 & 5.944 & 0.312 & 1.156 & 0.524 \\
2018 & 4.540 & 0.475 & 0.525 & 0.757 \\
2017 & 4.666 & 0.458 & 0.559 & 0.740 \\
2016 & 0.940 & 0.967 & 0.053 & 0.995 \\
\hline
\end{tabular}
\footnotesize
\item \textit{Notes}: The LM (Lagrange Multiplier) and F-statistics test for heteroskedasticity in the residuals of regression models. A $p$-value greater than 0.05 suggests failure to reject the null hypothesis of homoskedasticity. Across all years from 2016 to 2022, results indicate no significant evidence of heteroskedasticity at the 5\% significance level.
\end{center}
\end{table}

\begin{figure}[!hb]
  \centering \includegraphics[width=0.9\textwidth]{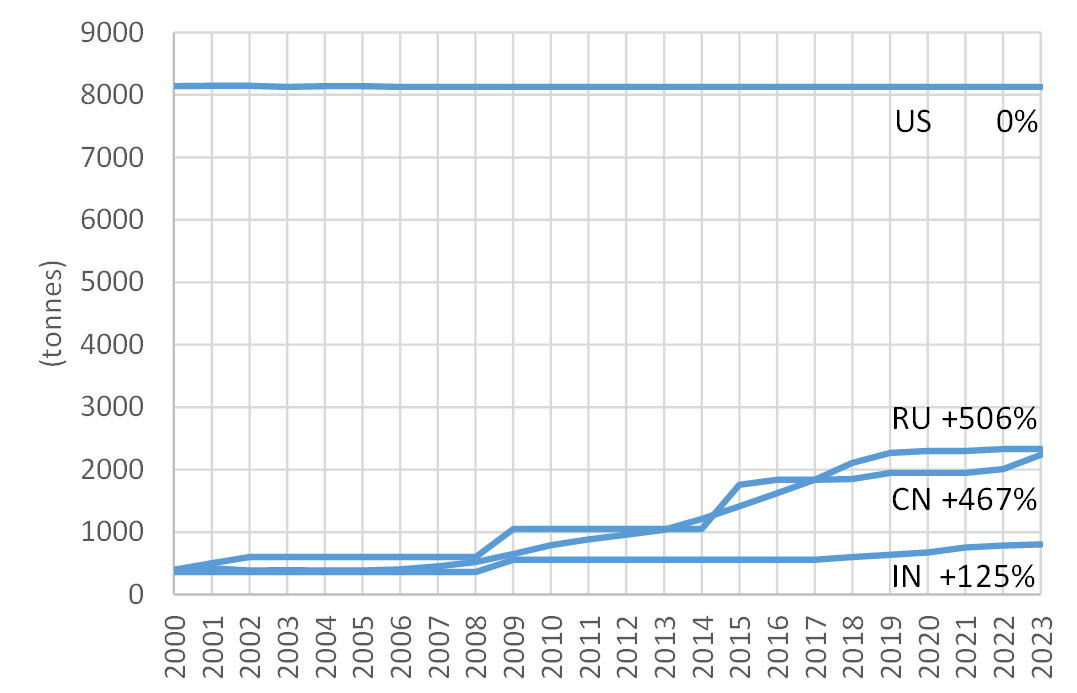}\\
    \caption{ Deposit of gold  in a set of BRICS countries, together with the U.S.}
  \label{gold1}
\end{figure}

\begin{figure}[!hb]
  \centering \includegraphics[width=0.9\textwidth]{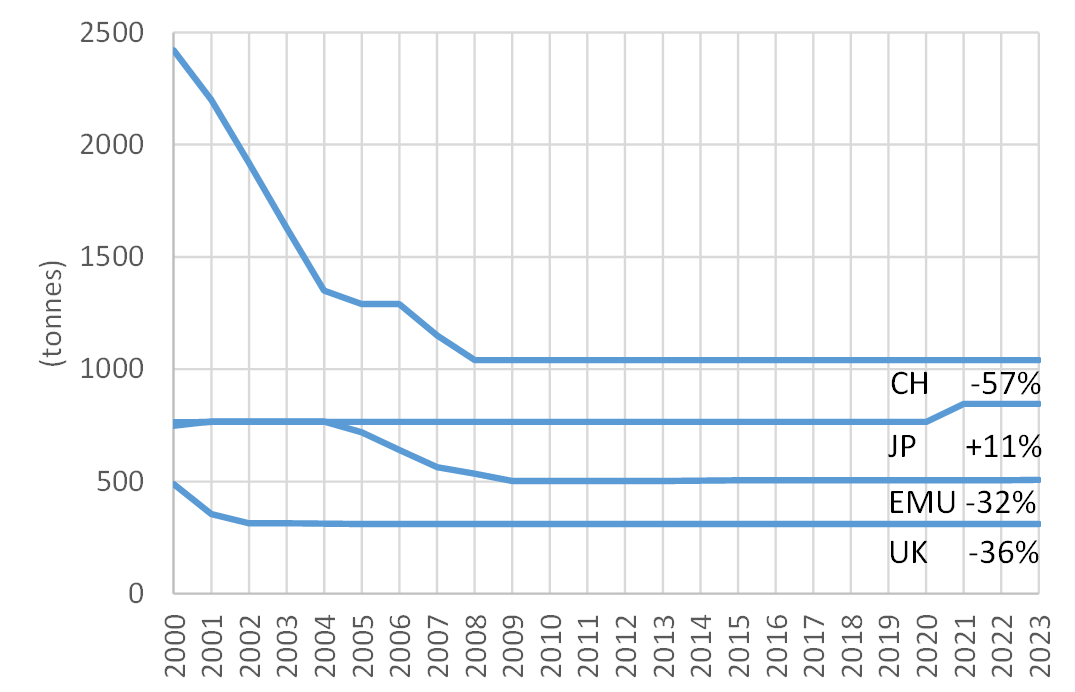}\\
  \caption{Deposit of gold in a set of developed countries. }
  \label{gold2}
\end{figure}

We again report that for each year separately between 2016 and 2022 the country's currency enlisted in Foreign Exchange Reserves significantly depends on the country's Military Spending (MS) chosen to represent the country's size. The estimated highly statistically significant coefficient on military spending is in line with the Mars hypothesis. Among controlling variables, gold reserves significantly negatively depend on FER for some years, where it is not surprising that the coefficient is negative, since gold and money are competitors, which is vividly shown for Russia and China which dollars have been replacing with gold.  Note that in contrast to China and Russia, developed western countries have decreased gold reserves during the last two decades. Table \ref{t6} shows that international trade is generally not a significant variable. Table \ref{t6} shows that FER significantly depends on the monetary instrument M2 for most of the years chosen (shown in Figs. \ref{gold1} and \ref{gold2}). However, for each year shown, MS is more statistically significant than M2. Comparing Tables \ref{t4} and \ref{t6}, we note that over time the inclusion of controlling variables reduces the coefficient
$ \beta_0$ much faster than without controlling variables. Table \ref{t7} shows that standard errors are robust to heteroskedasticity.

Extrapolating the data on the RMB dummy coefficient in Table \ref{t6}, when control variables are included, one can roughly predict that the RMB could reach an equilibrium with currencies from western countries in $ \approx  40 $ years.  Interestingly, when the M2 variable is omitted, we find that the RMB could reach an equilibrium in $ \approx  15$ years, a value closer to the one we obtained based on the data in Table \ref{t2}.

We also show regression analyses with a reduced set of controlling variables. Table \ref{t6} reveals that when M2 is omitted, the coefficient estimates are more affected than when the IT variables are omitted.

\subsection{Model} 

In globalization, China has probably benefited the most, and since approximately 1994, when China began to open up, its economic growth has been incredibly fast, in a way comparable to the growth of America in the nineteenth century. As a country grows economically, especially through international trade, more and more countries are using its currency. As China's GDP approaches the US GDP, especially measured in the PPP, the probability of geopolitical motives such as trade wars and tariffs starts to increase because the country that introduced globalization will not allow another country to overtake its primacy as a world economic leader. Namely, a country that would achieve the same GDP and then even larger GDP than the U.S:, especially in PPP terms,  would be in position to invest more money than the U.S. in the military, research and science, and as time passes, then that country would naturally become a new world leader. As we stated in the introductory part of the paper, there is a reason why there is no place on Earth where tigers and lions live together because they are predators of the same level. 

Similarly as in the living world where if lions and tigers met randomly, they would tear each other apart, in geopolitics, in order to preserve its primacy, the most powerful country in the world imposes geopolitical tools through sanctions and tariffs against the rising competing rival.  However,  as a countermeasure, the competing country also uses geopolitical tools and even initiates geopolitical tools in an economic war. If a particular country is a competitor, it will not help the economy of different geopolitical blocs by saving in their currency and using it in trades with other countries.  The introduction of tariffs and sanctions is an example of different geopolitical tools to increase country`s geopolitical position.  

We model the global monetary system as an undirected network \( G = (V,E) \) consisting of \( N=195 \) sovereign states, represented as nodes. Each edge \((i,j)\in E\) denotes economic interactions: trade flows, financial interdependencies, or strategic partnerships. Node \(0\) is assigned to the United States (USD) while nodes 1 and 2 represent a set of geopolitical rivals, China and Russia, explicitly assigned their own currencies (CNY, RUB) and economic growth characteristics. 
 The remaining nodes (\(3\)–\(194\)) represent independent states, initially dollarized but subject to network contagion-driven dedollarization. Each country \(i\) maintains three dynamic state variables: a GDP time series \(\mathrm{GDP}_i(m)\), a dedollarization level \(d_i(m)\in[0,1]\), and a primary currency label \(c_i(m)\in\{\mathrm{USD}, \mathrm{CNY}, \mathrm{RUB}\}\). 
We initialize GDP trajectories using historical annual data (2010 to the latest available year) for the United States, China, and Russia, which we interpolate to a monthly frequency.  Beyond empirical observations, each core series is extrapolated using fixed annual growth rates: 2\% for the U.S., 3\% for China, and 1\% for Russia, converted to equivalent monthly growth factors.  For peripheral states, we draw the initial GDPs from a Pareto distribution with shape parameter \(\alpha=2.5\) and scale \(x_{\min}=0.5\times10^{12}\), a choice motivated by the heavy-tailed heterogeneous size distribution observed in real‑world country GDPs, and then apply a uniform annual growth rate 2\% (also converted to a monthly factor).
The network employs a hub-and-spoke topology with the U.S.\ (node 0) at its center. Each non-U.S.\ node connects directly to the U.S.\ Additional peer-to-peer edges are introduced among non-U.S.\ countries to achieve a minimum degree of ten, with probabilities proportional to their GDP. Thus, larger economies exhibit greater centrality. We have found that GDP and military spending correlate with a surprisingly high $R^2$, suggesting that GDP serves as an excellent proxy for military spending. In other words, based on the analysis of \citet{clements2021military}, there is a general convergence to a steady state of military spending as a share of GDP between countries. Therefore, implicitly in our analysis, GDP growth also represents a proportionally growth of military spending. Based on this finding, we assume in the model that GDP and military spending are linearly dependent, such that $\text{MS} = \gamma \cdot \text{GDP}$. Thus, in our model, GDP effectively serves as a proxy for military spending.

\begin{figure}[!htbp]
     \caption{Yearly U.S. imports of goods from China, reported on a customs basis in millions of U.S. dollars.}
  \centering \includegraphics[width=0.99\textwidth]{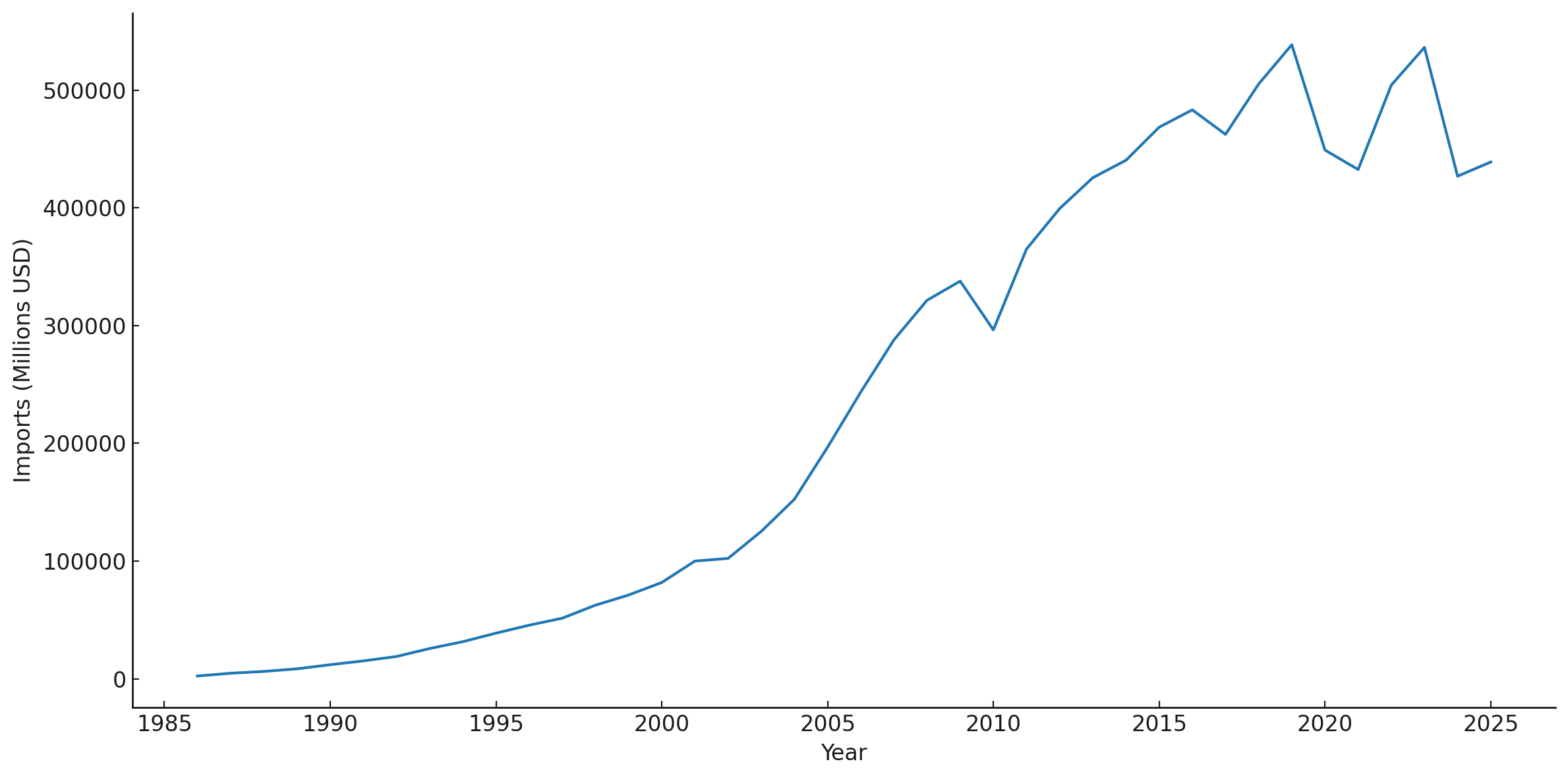}

  \label{import}
\end{figure}

Precisely,  we model the international economic system started with the fall of socialism when globalization gained its momentum and many countries including China began their rapid growth. During the period of globalization, the world FER market was in a state of equilibrium or at least quasi-equilibrium,  where only the currencies of developed Western democracies were being used for the savings of all countries.  
However, in the model, we hypothesize that \textit{as China approaches the US in terms of GDP, the probability that the U.S. will respond with geopolitical moves such as sanctions, reduction of imports from China, ban of chip exports, and tariffs ramps up, where as a response China automatically responds with de-dollarization.}  In agreement with this hypothesis, Fig. \ref{import} shows how USA imports from China grew continuously for years until China reached the American GDP in terms of purchasing power. 
We further hypothesize that \textit{if the public debt interest of the global leader exceeds its military spending, it leads to a declining confidence in the global leader currency } in agreement with Ferguson's law which states that any great power that spends more on debt than on military spending risks ceasing to be a great power.

We set the constraint that  the U.S., as a global leader, wants to preserve its economic dominance and is eager to respond using economic terms  if the country's GDP of any US rival approaches the U.S. GDP. Therefore, a dedollarization cascade begins at the time \( t^* \), when China’s GDP first surpasses that of the U.S. Prior to this point, independent nodes are fully dollarized (\(d_i(t)=0\)). In addition, as we discovered in the empirical part of the research, dedollarization is perhaps an inappropriate name for the financial process in which the FER market mainly due to Chinese RMB simply moves from one equilibrium level to the next one. We model  this tendency towards a new FER equilibrium by postulating that for simplicity reason, after \(t^*\), dedollarization pressures initiate with maximum intensity, linearly decaying over a horizon of \(L=180\) months:
\[
\delta(t) = \max\left(0,\,1 - \frac{\max(0,t-t^*)}{L}\right).
\]
where in agreement with our findings, it takes 15 years (or 180 months) to reach a new equilibrium. Two contagion mechanisms operate simultaneously after \(t^*\):

\paragraph{Inherent Shocks:} In the model, when China’s GDP first surpasses U.S. GDP at time $t^*$, the U.S. triggers sanctions and imposes restrictions on the U.S. market (see Fig. \ref{import}). In response, China initiates dedollarization (see Fig. \ref{CHNUS}). Each node independently undergoes dedollarization increments with probability $p_0(t) = 0.02\,\delta(t)$, representing idiosyncratic economic shocks or policy decisions.

\paragraph{Neighbor Influence:} Nodes are influenced by their economic neighbors (excluding the U.S.). Let $\bar{d}(t)$ represent the average level of dedollarization of the neighbors of node $i$. If $\bar{d}(t) \geq 0.25$ and the US debt service obligations exceed its military spending, node $i$ experiences a similar random dedollarization increase with probability $p(t) = 0.008\,\delta(t)$. In this context, the military power of the U.S., as reflected in its fiscal capacity, affects the willingness of other countries to move away from the dollar in their savings and possibly in trade.\\

When the dedollarization of a node exceeds \(0.5\) and remains labeled USD, it adopts the most common alternative currency (CNY or RUB) among its neighbors. In bonds, assignment occurs randomly with equal probability. Finally, after 180 steps, the system reaches a new FER equilibrium. 

In order to explore the predictive power  of the network model for instance  where we could expect a new FER equilibrium in the future, for simplicity's sake, let's assume that at the beginning of simulations each country saves equally in dollars, with fraction say 0.56. As we have information about GDP for each country i, it makes sense to assume that savings in dollars are some fraction of GDP, say $FER_i = \beta GDP_i.$  For a given set of network parameters, at each step we know how many countries, accepted dedollarization either inherently or externally. In the model, when a country inherently or externally accepts dedollarization, it is randomly assigned the extent to which it accepts dedollarization, where it is any fraction between 0 and the FER fraction of dollar savings $f_t$ in that step. Then at the following step we  calculate the time dependent  fraction of FER in dollars $f_{t+1}$.

Each simulation step includes a monetary expansion event, wherein the United States injects a fixed quantity of dollars (\( M_{\text{printed}} \)) into the global economy. These dollars are distributed among dollarized countries proportionally to their GDP, simulating the global spread of US liquidity through trade and finance. The effective inflation rate faced by the United States is computed as:

\[
I_{\text{eff}} = \frac{I_{\text{base}}}{1 - f}
\]

where \( f \) denotes the fraction of countries that are partially or fully dedollarized. As \( f \rightarrow 1 \), inflation rises asymptotically, simulating the collapse of global dollar demand. The interest payments on US public debt increase proportionally with monetary expansion, adding fiscal stress and feedback into the external de-dollarization mechanism. This interaction captures a potential tipping point, where fiscal irresponsibility may directly undermine global trust in the USD.
 
To assess the robustness of our dedollarization framework under stochastic variability, we perform \(R=100\) independent Monte Carlo runs. Each run generates time series for the global dedollarization fraction, the effective U.S.\ inflation rate, and the currency‐share per country. From these values we compute ensemble averages. To analyze system evolution, the model tracks several macroeconomic indicators over time, including:

\begin{itemize}
    \item The average dedollarization score across all countries.
    \item The effective US inflation rate.
    \item The distribution of global currencies (USD, CNY, RUB).
    \item US public debt interest burden.
    \item The volume of dollars received by each country.
\end{itemize}

Snapshots of the network state are captured at multiple intervals to visualize structural transitions, such as the emergence of new currency blocs or the reduction of US-centric connectivity.

\subsection{Simulation Results} 

\begin{figure}[!htbp]
     \caption{The spread of de-dollarization (top) and its impact on US inflation (bottom). As de-dollarization increases over time, US inflation also rises, indicating an economic feedback loop.}
  \centering \includegraphics[width=0.8\textwidth]{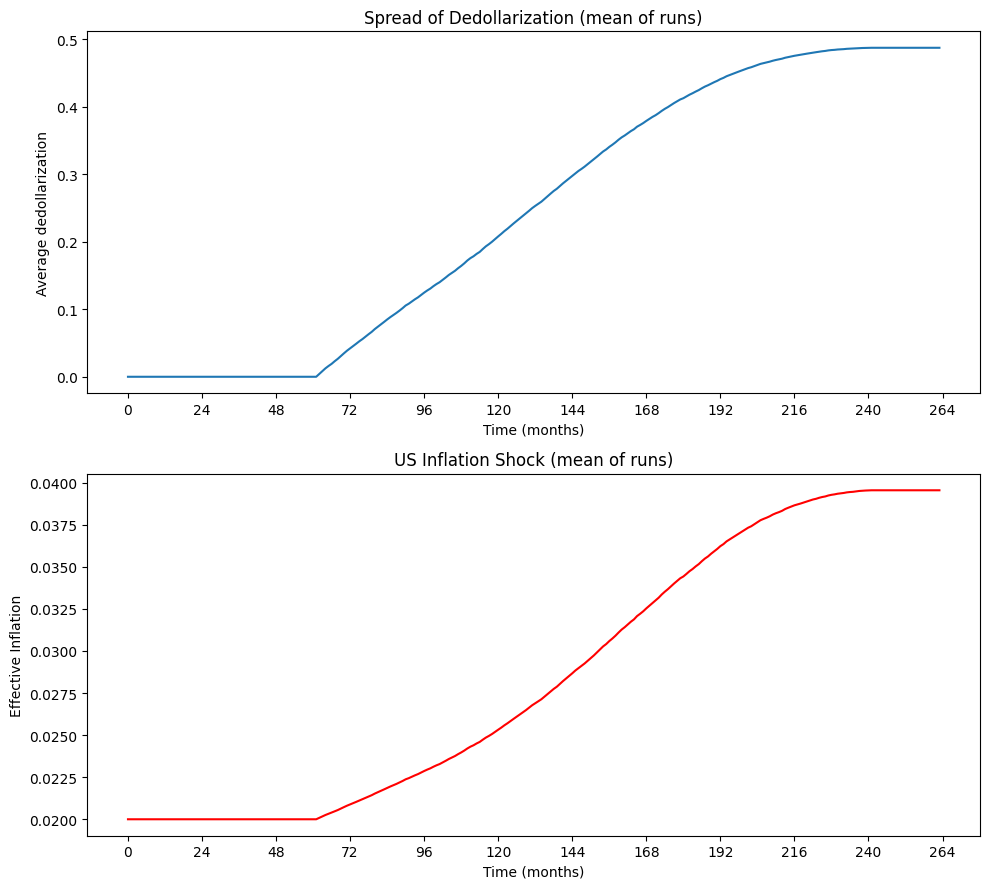}\\

  \label{dedol_infl}
\end{figure}

The simulation’s ensemble‐mean dedollarization trajectory (Figure \ref{dedol_infl}) exhibits three distinct regimes. In the first 60 time steps, the average dedollarization fraction remains at zero, reflecting the period prior to China’s GDP overtaking that of the United States and before any contagion begins. Once the cascade is triggered, dedollarization spreads in an approximately linear fashion, climbing from zero to roughly 0.25 in the first 130 time steps. Thereafter the propagation accelerates slightly, reaching around 0.35 by step 150 and approaching 0.5 by step 220. This shape reflects the time‐decaying contagion probability---which dampens new shocks over a 15‑year horizon---and the cumulative effect of peer influence once a critical mass of neighbors has moved away from USD.

The corresponding inflation shock remains pinned at the base rate of 2\% until the dedollarization cascade commences. Thereafter, effective U.S.\ inflation rises gradually—first to about 2.2\% by step 100 and then accelerating to roughly 3\% by step 150—before climbing more steeply to around 4\% by step 220. This convex profile arises because inflation is modeled as

\[
I_{\mathrm{eff}}\;=\;\frac{I_{\mathrm{base}}}{1 - f}\;,
\]

so that as the dedollarization fraction \(f\) grows, marginal increases in \(f\) exert an ever‑larger inflationary impact. The late‑horizon uptick reflects both the saturation of dedollarization—and hence the divergence of the denominator—and the cumulative build‐up of U.S.\ debt‑service costs, which tightens the neighbor‐influence channel and sustains upward pressure on \(\pi\).

\begin{figure}[!htbp]
     \caption{Ensemble‐mean evolution of primary currency shares. Following the dedollarization cascade onset, the USD’s share (blue) declines steadily from near‐unity to approximately 50 percent, while the CNY (orange) rises to about 40 percent and the RUB (green) to roughly 10 percent, signaling the emergence of a multipolar reserve‐currency regime.}
  \centering \includegraphics[width=0.99\textwidth]{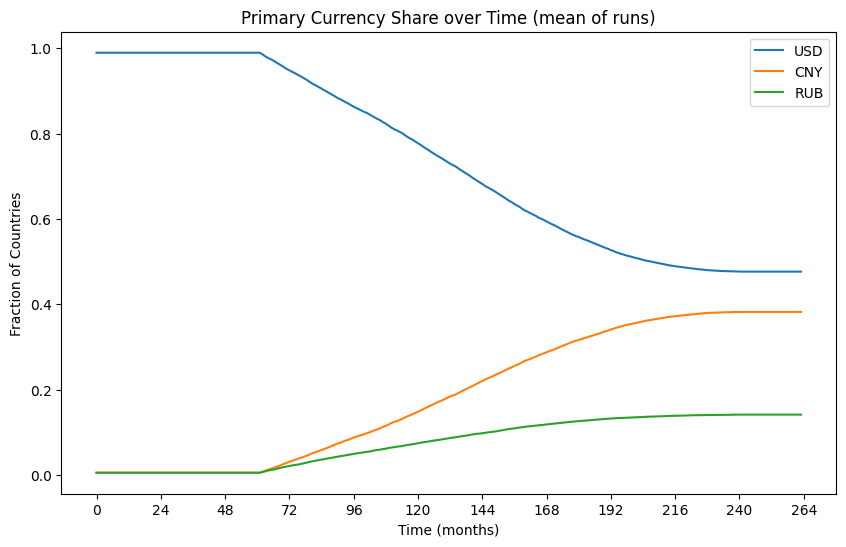}

  \label{curr}
\end{figure}

In the currency‐share series (Figure \ref{curr}), the USD share begins at nearly 100\% (1.0 on the vertical axis), then declines steadily, falling to approximately 75\% by step 150, 60\% by 2022, and stabilizing around 50\% by 2028. The CNY emerges as the principal alternative, rising to about 10\% share by step 100, 25\% by step 180, and reaching roughly 40\% by step 220. The RUB share grows more modestly—to about 10\% by step 150—reflecting its more limited peer‐to‐peer influence and lower initial network centrality. Together these trajectories illustrate a multipolar currency regime in which China’s renminbi captures the lion’s share of dedollarized economies, while the ruble plays a secondary regional role.

\begin{figure}[!htbp]
     \caption{Network snapshot illustrating the micro‑state of dedollarization contagion. Gray nodes remain USD‑pegged, red nodes have switched to CNY, and blue nodes to RUB. The U.S. hub (node 0) retains high centrality, while CNY and RUB adopters form contiguous clusters around China (node 1) and Russia (node 2), respectively.}
  \centering \includegraphics[width=0.99\textwidth]{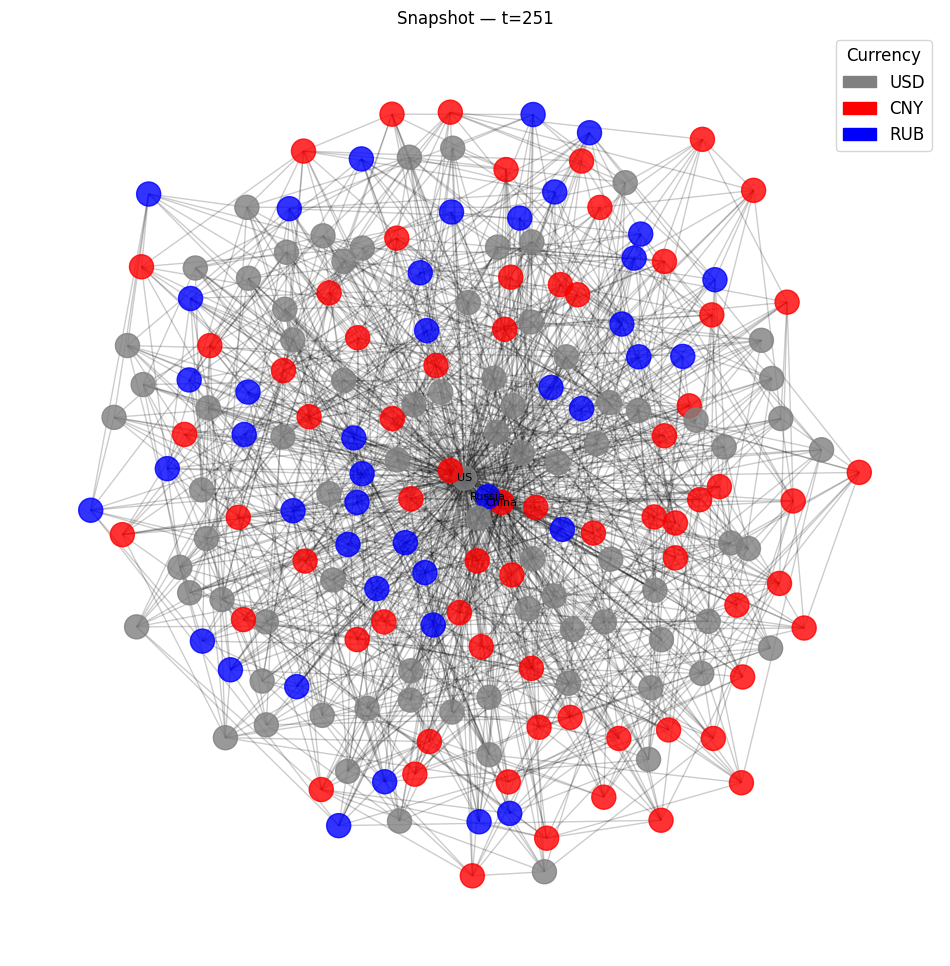}\\

  \label{network}
\end{figure}

A snapshot of the network (Figure \ref{network}) makes these aggregate patterns concrete at the micro‐level. Node 0 (the U.S.) remains an entrenched hub, but roughly half of the peripheral nodes have switched to CNY (red) and a smaller cluster to RUB (blue), leaving the remainder still pegged to USD (gray). CNY‐adopting countries tend to cluster around node 1 (China) and among each other, forming contiguous red communities that show GDP‐weighted attachment and neighbor‐influence dynamics. RUB adopters likewise form local blue clusters around node 2 (Russia) and along Russian trade corridors. Gray nodes are increasingly isolated, connected mainly to the U.S.\ hub but lacking sufficient neighbor pressure to switch. The dense web of black edges makes partially dedollarized states remain structurally embedded in the broader network, suggesting that full fragmentation of U.S.\ centrality would require deeper and more widespread currency realignments.

\begin{figure}[!htbp]
     \caption{USD‑only subgraph (December 2027): Approximately half of all states remain dollar‑pegged, forming a dense hub‑and‑spoke network with the U.S. at its center.}
  \centering \includegraphics[width=0.99\textwidth]{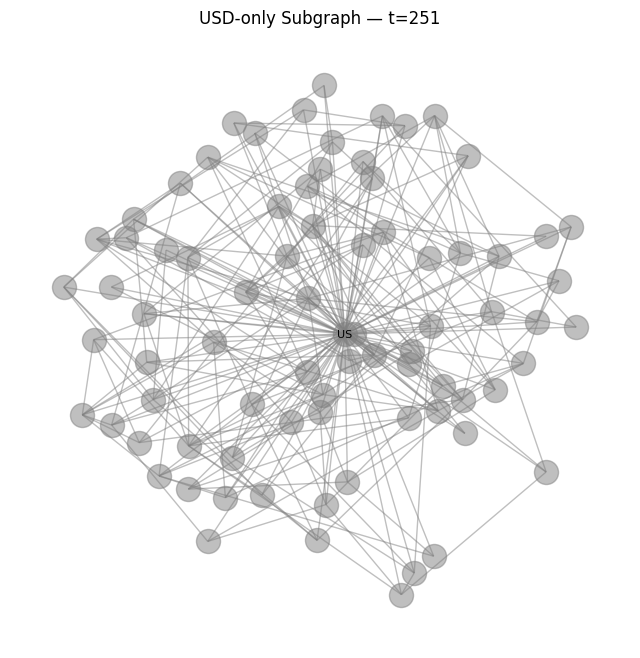}\\

  \label{us_network}
\end{figure}

In the USD‑only subgraph (Figure \ref{us_network}) we see roughly half of all countries still pegged to the dollar. This cluster remains extremely dense—most nodes are directly linked to the U.S. hub (node 0) and, thanks to the original “hub‑and‑spoke” plus peer‑attachment construction, many of those peripheral nodes also maintain significant interconnections amongst themselves. Visually, the U.S. node sits dead‑center with a spider‑web of gray edges; the high local clustering and large average degree indicate that dollarized states still form a highly cohesive core. Even as other blocs emerge, these gray‑pegged countries continue to enjoy robust mutual linkages.

\begin{figure}[!htbp]
     \caption{CNY‑only subgraph: The largest non‑USD community—centered on China (node 1). Although sparser than the USD network, it exhibits a robust clustering pattern, with many red nodes linked both to China and to one other.}
  \centering \includegraphics[width=0.99\textwidth]{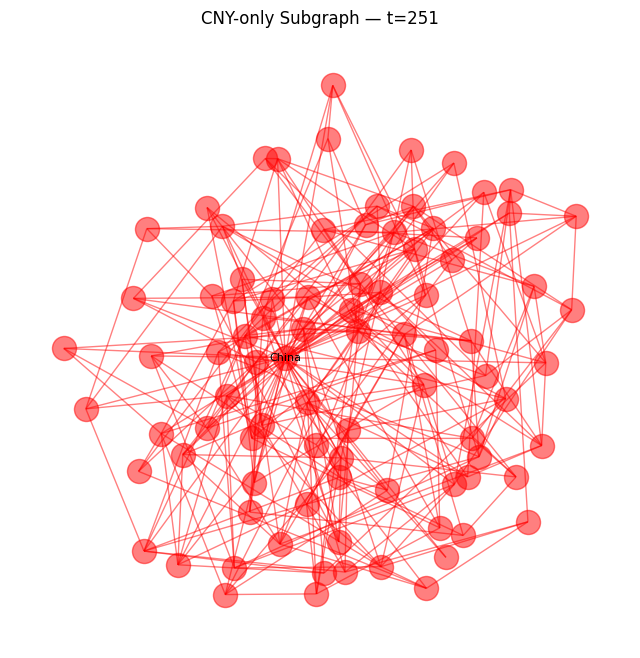}\\

  \label{china_network}
\end{figure}

The CNY‑only network (Figure \ref{china_network}) comprises the largest non‑USD community. Compared to USD, this red cluster is somewhat sparser but still shows clear cohesion: many red nodes share edges with both China and one another, forming multiple overlapping triads. The elevated clustering coefficient here reflects regional or trade‐corridor effects: once a country switches to CNY, it tends to bring along several of its neighbors. The CNY cluster’s fairly uniform radial spread around node 1 (China) signals that renminbi adopters are not just direct Chinese trading partners but also peers of peers, creating a secondary core in the global network.

\begin{figure}[!htbp]
     \caption{RUB‑only subgraph: A small, regionally concentrated cluster surrounds Russia (node 2). Low average degree and minimal clustering beyond Russia’s immediate neighbors indicate that the ruble’s sphere of influence remains limited to a tight inner circle rather than a broad global core.}
  \centering \includegraphics[width=0.99\textwidth]{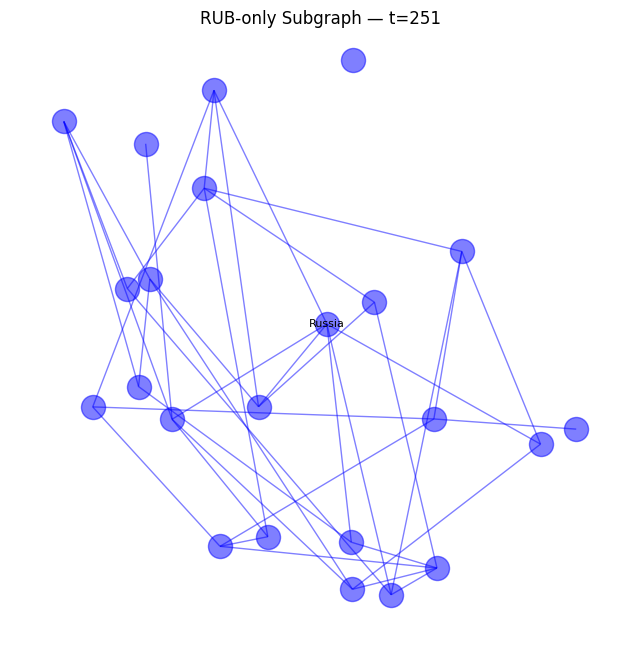}\\

  \label{russia_network}
\end{figure}

By contrast, the RUB‑only network (Figure \ref{russia_network}) is much smaller—only a couple dozen nodes—and considerably more fragmented. Node 2 (Russia) serves as the focal point, yet many blue nodes lie at distance two or three, connected by only one or two edges. The average degree here is low, and the clustering coefficient is nearly zero outside Russia’s immediate neighbors, indicating only small regional clusters rather than a broad‐based community. In effect, the ruble’s sphere remains limited: it captures a tight “inner circle” of Russian‐aligned economies, but beyond that, adoption is too sparse to form a full structural core.

\section{Discussion. }  

Military spending serves not just to protect the country's borders, but in the case of the largest countries to protect the country's economic interests and vouch for stability when other countries are investing in its currency. In globalization, every nation competes with all others, and if it is more competitive, it has a chance to develop faster than  the  less competitive  ones. Throughout the history we have seen how changes in military and economic power are interconnected, hence the importance of a leader`s currency in the world trade. The UK's pound sterling was the primary reserve currency of the world in the 19th century and the first half of the 20th century. However, by the mid 20th century, the United States dollar had overtook pound sterling as the leading world`s currency. 

Currently, the dollar's dominance as an international reserve currency correlates with the U.S. global military position and its role to guarantee the security of its allies. 
In this paper, inspired by Mars hypothesis \citet{eichengreen2019mars}  used to explain composition of FER, we report how the country's GDP and the country's MS highly correlates with the country's currency enlisted in Foreign Exchange Reserves (FER) ($R^2 $ always higher than 0.95), for a set of Western countries all characterized by liberal democracy, for each year over the past 20 years, supporting the hypothesis that these countries are presumably in an equilibrium state involving GDP, military spending and FER composition. Moreover, military spending explains slightly better than GDP this type of equilibrium.

However, in aforementioned equilibrium the Chinese RMB and the overall set of other currencies can be considered as outliers pushing the FER composition in the direction of a new equilibrium where the RMB will take the larger share. The Chinese RMB is currently well beyond the FER balance, but there is a steady trend towards a new FER balance.  By assuming geopolitical relationships similar to those that have been present in the last 20 years, our analysis indicates a trend such that RMB could reach a new FER balance within 15 years. However, our prediction is just an extrapolation of past behavior, and we are perfectly aware that predictions on the FER currency composition have been shown to be wrong \citet{chinn2008euro}. In our network model changes to a different FER currency are mechanically connected to international trade. As certain country`s  share of international trade grows at an above average rate, its closest neighbors will trade more in its currency, so the growth of dedollarization in our model is implicitly related to the growth of international trade.  Empirically, that was a significant feature of China`s growth over the past few decades.
   
In order to accept the RMB as a new alternative reserve currency, many experts believe that China should first substantially expand access
to its financial market and, second, increase imports the same as the US did in order to offer dollars as a dominant reserve currency. Recently, the Chinese premier Li Qiang told in Shanghai during the annual China International Import Expo that China was committed to opening up its economy, where he speculated that Chinese imports of goods and services were predicted to be around 17 trillion dollars in the next five years. The potentially larger share of RMB in the FER pool implies the smaller share of other currencies.

In addition to a gradual diversification of international reserves, an alternative scenario can lead to a partial or complete financial and economic breakdown between the United States and China according to \citet{eichengreen2024international}. In our model, when China’s GDP surpasses that of the U.S., the U.S. reacts by imposing sanctions and tightening access to its domestic market. In response, China accelerates dedollarization efforts, a dynamic consistent with the patterns observed in recent empirical data. Data cannot be used to predict the possible outcome of the world economy, but our model results suggest that we are moving towards a new global multi-currency FER equilibrium, where the dedolarization level will depend not only on economic factors, but also on geopolitical factors and how US will be able to maintain good geopolitical ties with their allies, old and new.

\newpage


\bibliographystyle{plainnat}
\bibliography{references}

\begin{thebibliography}{30}
\providecommand{\natexlab}[1]{#1}
\providecommand{\url}[1]{\texttt{#1}}
\expandafter\ifx\csname urlstyle\endcsname\relax
  \providecommand{\doi}[1]{doi: #1}\else
  \providecommand{\doi}{doi: \begingroup \urlstyle{rm}\Url}\fi

\bibitem[imf(2024)]{imfdata2024}
International monetary fund data portal.
\newblock \url{https://data.imf.org/?sk=e6a5f467-c14b-4aa8-9f6d-5a09ec4e62a4}, 2024.
\newblock Accessed April 2025.

\bibitem[Ahmed et~al.(2023)Ahmed, Aizenman, Saadaoui, and Uddin]{ahmed2023effectiveness}
Rashad Ahmed, Joshua Aizenman, Jamel Saadaoui, and Gazi~Salah Uddin.
\newblock On the effectiveness of foreign exchange reserves during the 2021-22 us monetary tightening cycle.
\newblock \emph{Economics Letters}, 233:\penalty0 111367, 2023.

\bibitem[Alwadeai et~al.(2024)Alwadeai, Vlasova, Mareeh, and Aljonaid]{alwadeai2024beyond}
Ahmed Alwadeai, Nataliia Vlasova, Hadi Mareeh, and Nadeem Aljonaid.
\newblock Beyond traditional defenses: Unraveling the dynamics of reserves and exchange rate volatility in the face of economic sanctions.
\newblock \emph{Russian Journal of Economics}, 10\penalty0 (1):\penalty0 1--19, 2024.

\bibitem[Arslan and Cant{\'u}(2019)]{arslan2019size}
Yavuz Arslan and Carlos Cant{\'u}.
\newblock The size of foreign exchange reserves.
\newblock \emph{BIS paper}, \penalty0 (104a), 2019.

\bibitem[Barnum et~al.(2025)Barnum, Fariss, Markowitz, and Morales]{barnum2025measuring}
Miriam Barnum, Christopher~J Fariss, Jonathan~N Markowitz, and Gaea Morales.
\newblock Measuring arms: Introducing the global military spending dataset.
\newblock \emph{Journal of Conflict Resolution}, 69\penalty0 (2-3):\penalty0 540--567, 2025.

\bibitem[Chinn and Frankel(2008)]{chinn2008euro}
Menzie Chinn and Jeffrey Frankel.
\newblock The euro may over the next 15 years surpass the dollar as the leading international currency.
\newblock Working Paper 13909, National Bureau of Economic Research, 2008.
\newblock Available at \url{http://www.nber.org/papers/w13909}.

\bibitem[Chitu et~al.(2014)Chitu, Eichengreen, and Mehl]{chitu2014dollar}
Livia Chitu, Barry Eichengreen, and Arnaud Mehl.
\newblock When did the dollar overtake sterling as the leading international currency: Evidence from the bond markets.
\newblock \emph{Journal of Development Economics}, 111:\penalty0 225--245, 2014.

\bibitem[Clements et~al.(2021)Clements, Gupta, and Khamidova]{clements2021military}
B~Clements, S~Gupta, and S~Khamidova.
\newblock Military spending in the post-pandemic era: countries’ efforts to secure a more peaceful world could have a positive economic effect.
\newblock \emph{Finance \& Development}, 2021.

\bibitem[Devereux and Shi(2013)]{devereux2013vehicle}
Michael Devereux and Shouyong Shi.
\newblock Vehicle currency.
\newblock \emph{International Economic Review}, 54:\penalty0 97--133, 2013.

\bibitem[Dominguez et~al.(2012)Dominguez, Hashimoto, and Ito]{dominguez2012international}
Kathryn~ME Dominguez, Yuko Hashimoto, and Takatoshi Ito.
\newblock International reserves and the global financial crisis.
\newblock \emph{Journal of International Economics}, 88\penalty0 (2):\penalty0 388--406, 2012.

\bibitem[Eichengreen(2024)]{eichengreen2024international}
Barry Eichengreen.
\newblock International finance and geopolitics.
\newblock \emph{Asian Economic Policy Review}, 19:\penalty0 84--100, 2024.

\bibitem[Eichengreen and Flandreau(2008)]{eichengreen2008rise}
Barry Eichengreen and Marc Flandreau.
\newblock The rise and fall of the dollar, or when did the dollar replace sterling as the leading reserve currency?
\newblock Working Paper 14154, National Bureau of Economic Research, 2008.

\bibitem[Eichengreen and Sachs(1985)]{eichengreen1985exchange}
Barry Eichengreen and Jeffrey Sachs.
\newblock Exchange rates and economic recovery in the 1930s.
\newblock \emph{Journal of Economic History}, 45:\penalty0 925--946, 1985.

\bibitem[Eichengreen and Sachs(1986)]{eichengreen1986competitive}
Barry Eichengreen and Jeffrey Sachs.
\newblock Competitive devaluation and the great depression: A theoretical reassessment.
\newblock \emph{Economics Letters}, 22:\penalty0 67--71, 1986.

\bibitem[Eichengreen et~al.(2019)Eichengreen, Mehl, and Chi{\c{t}}u]{eichengreen2019mars}
Barry Eichengreen, Arnaud Mehl, and Livia Chi{\c{t}}u.
\newblock Mars or mercury? the geopolitics of international currency choice.
\newblock \emph{Economic Policy}, 34\penalty0 (98):\penalty0 315--363, 2019.

\bibitem[Ito and McCauley(2020)]{ito2020currency}
Hiro Ito and Robert~N McCauley.
\newblock Currency composition of foreign exchange reserves.
\newblock \emph{Journal of International Money and Finance}, 102:\penalty0 102104, 2020.

\bibitem[James et~al.(2013)James, Witten, Hastie, Tibshirani, and Taylor]{james2013introduction}
Gareth James, Daniela Witten, Trevor Hastie, Robert Tibshirani, and Jonathan Taylor.
\newblock \emph{An Introduction to Statistical Learning}.
\newblock Springer Texts in Statistics. Springer New York, New York, NY, 2013.

\bibitem[Jeanne and Ranci{\`e}re(2011)]{jeanne2011optimal}
Olivier Jeanne and Romain Ranci{\`e}re.
\newblock The optimal level of international reserves for emerging market countries: A new formula and some applications.
\newblock \emph{The Economic Journal}, 121\penalty0 (555):\penalty0 905--930, 2011.

\bibitem[Kennedy(1989)]{kennedy1989rise}
Paul Kennedy.
\newblock \emph{The Rise and Fall of the Great Powers: Economic Change and Military Conflict from 1500 to 2000}.
\newblock Fontana, London, 1989.

\bibitem[King and Levine(1993)]{king1993finance}
Robert~G. King and Ross Levine.
\newblock Finance and growth: Schumpeter might be right.
\newblock \emph{Quarterly Journal of Economics}, 108:\penalty0 717--737, 1993.

\bibitem[Krugman(1980)]{krugman1980vehicle}
Paul Krugman.
\newblock Vehicle currencies and the structure of international exchange.
\newblock \emph{Journal of Money, Credit and Banking}, 12:\penalty0 513--526, 1980.

\bibitem[Levy-Yeyati and G{\'o}mez(2020)]{levy2020cost}
Eduardo Levy-Yeyati and Juan~Francisco G{\'o}mez.
\newblock \emph{The cost of holding foreign exchange reserves}.
\newblock Springer, 2020.

\bibitem[Li and Liu(2008)]{li2008rmb}
David Li and Linlin Liu.
\newblock Rmb internationalization: An empirical analysis.
\newblock \emph{Journal of Financial Research}, 11:\penalty0 1--16, 2008.

\bibitem[Matsuyama et~al.(1993)Matsuyama, Kiyotaki, and Matsui]{matsuyama1993toward}
Kiminori Matsuyama, Nobuhiro Kiyotaki, and Akihiko Matsui.
\newblock Toward a theory of international currency.
\newblock \emph{Review of Economic Studies}, 60:\penalty0 283--307, 1993.

\bibitem[Modelski and Thompson(1996)]{modelski1996leading}
George Modelski and William~R. Thompson.
\newblock \emph{Leading Sectors and World Powers: The Coevolution of Global Politics and Economics}.
\newblock Studies in International Relations. University of South Carolina Press, Columbia, SC, 1996.

\bibitem[Peltier(2023)]{peltier2023we}
Heidi Peltier.
\newblock We get what we pay for: the cycle of military spending, industry power, and economic dependence.
\newblock \emph{Costs of War}, 2023.

\bibitem[Portes and Rey(1998)]{portes1998euro}
Richard Portes and Hélène Rey.
\newblock The emergence of the euro as an international currency.
\newblock \emph{Economic Policy}, 13:\penalty0 307--343, 1998.

\bibitem[Rodrik(2006)]{rodrik2006social}
Dani Rodrik.
\newblock The social cost of foreign exchange reserves.
\newblock \emph{International economic journal}, 20\penalty0 (3):\penalty0 253--266, 2006.

\bibitem[Romer and Romer(2000)]{romer2000federal}
Christina~D. Romer and David~H. Romer.
\newblock Federal reserve information and the behavior of interest rates.
\newblock \emph{American Economic Review}, 90\penalty0 (3):\penalty0 429--457, 2000.

\bibitem[Triffin(1960)]{triffin1960gold}
Robert Triffin.
\newblock \emph{Gold and the Dollar Crisis}.
\newblock Yale University Press, New Haven, 1960.

\end{thebibliography}

\end{document}